# Cosmic-Ray Antiprotons From Neutralino Annihilation into Gluons

Gerard Jungman[†]

*Department of Physics, Syracuse University, Syracuse, NY 13244.*
*and*
*Enrico Fermi Institute, University of Chicago, Chicago, IL 60637*

and

Marc Kamionkowski[‡]

*School of Natural Sciences, Institute for Advanced Study, Princeton, NJ 08540*

## ABSTRACT

We estimate the flux of cosmic-ray antiprotons expected from the annihilation of neutralinos in the galactic halo. The antiproton signal may offer an important alternative detection scheme in the case that neutralino annihilation proceeds mainly to the two-gluon final state.

October 1993

[†] jungman@npac.syr.edu
[*] Current address
[‡] kamion@guinness.ias.edu

# 1. Introduction

Luminous matter almost certainly does not account for all matter in the Universe [1], and this matter deficit inspires both particle-physics and astrophysics speculation. Among the particle-physics solutions discussed, perhaps the most attractive idea is that stable weakly interacting massive particles (WIMPs) may make up the dark matter. Currently, the most promising candidate WIMP is the neutralino [2][3], a linear combination of the supersymmetric partners of the photon, $Z^0$, and Higgs bosons.

Recently, we have presented a complete calculation of the cross-section for neutralino annihilation to the two-gluon final state [4]. This one-loop process can be competitive with tree-level annihilation processes in large regions of the parameter space of the supersymmetric standard model. The main objective of Ref. [4] was to calculate the resulting dilution of the secondary-neutrino spectrum, which is of interest since such secondary neutrinos can provide a detectable signal for annihilation of neutralinos captured in the Sun or the Earth [5]. A large branching ratio to the two-gluon final state implies a decreased branching ratio for the annihilation to heavy-fermion final states, which produce the signal secondary-neutrino spectrum via weak decay.

In Ref. [4], the possibility was discussed that the gluonic final state, though decreasing the detectability via secondary neutrinos, may open another avenue of detection—low-energy cosmic-ray antiprotons from WIMP annihilation in the galactic halo [6]. Hadronization of the final-state gluon jets will produce antiprotons in some fraction of the annihilation events. Although the "background" spectrum of cosmic-ray antiprotons produced by interactions of primary cosmic rays with interstellar matter is quite uncertain, the antiproton spectrum in standard cosmic-ray propagation models falls steeply at energies less than about a GeV [7]. If neutralinos populate the halo, the resulting distortion of the background spectrum due to antiprotons from WIMP annihilation may be significant at low energies. Previous work has considered the possibility that a detectable cosmic-ray antiproton signal could arise from the hadronization of b-quark and c-quark jets in the tree-level annihilation processes [8][9], and we must include this mode for antiproton production as well, though our discussion will concentrate on the new gluonic contribution.

Due to numerous astrophysical uncertainties, it is impossible to make a precise prediction for the antiproton spectrum for a given particle-physics model. However, we can show, for reasonable astrophysical assumptions, that a low-energy cosmic-ray antiproton signal from WIMP annihilation may indeed be observable.

In the following we will give an estimate of the source spectrum for antiprotons produced by neutralino annihilation. Then we will present numerical results for the antiproton spectrum at the Earth, for some model parameters. We concentrate on a simple $B$-ino WIMP scenario to illustrate the possibilities.

# 2. Source Function for Cosmic-Ray Antiprotons

In the annihilation of neutralinos to two gluons, $\tilde{\chi}\tilde{\chi} \to gg$, the back-to-back gluonic jets, each with energy $m_{\tilde{\chi}}$, will shower to produce color-singlet hadronic states, notably pions, kaons, (anti)protons, and (anti)neutrons. This process is described by fragmentation functions, $D_h^{gg}(z)$, one for each species of hadron, $h$, which give the distribution in energy of the final-state hadrons as a fraction of the initial jet energy, $z = E_{\text{had}}/E_{\text{jet}}$. There is also a small logarithmic dependence on the jet energy, calculable in perturbative QCD [10]. Clearly the spectrum of injected antiprotons is precisely (twice) the fragmentation spectrum for the hadronization of a gluon jet, $g \to p\bar{p} + X$. Therefore we must assemble what is known about gluonic jet fragmentation to (anti)protons. Note that we are interested in the properties of gluon jets with energies below about 150 GeV, since the branching ratio to the gluonic final state can be appreciable only when the neutralino mass is smaller than the top-quark mass and, in some cases, the $W$- and $Z$-boson masses.

The fragmentation process is an inherently non-perturbative QCD process, beyond our ability to calculate at the present time. Therefore, we rely on the measured properties of QCD jets and on fragmentation models when necessary. Unfortunately, very few properties of gluon jets are measured well. Though gluon jets are produced copiously in $p\bar{p}$ collisions, the required exclusive analysis of the final state in such collisions is extremely difficult. Many properties of quark jets have been measured in $e^+e^-$ collisions [11][12][13], but the analysis of three-jet events, attributed to the final state $q\bar{q}g$, has not produced the required detailed gluon fragmentation information. However, a combination of theoretical arguments and experimental information can be used to relate the properties of gluon and quark jets. Furthermore, detailed Monte-Carlo fragmentation models provide some insight into gluonic jets; the reliability of such models is judged based on their ability to predict many aspects of jet physics in $e^+e^-$ collisions [12].

As an empirical fact, the fragmentation functions for jets produced in $e^+e^-$ annihilation can be parametrized in the form

$$\frac{1}{\sigma}\frac{d\sigma(e^+e^- \to p\bar{p}X)}{dz} = \beta C \sum_i a_i \exp{-b_i z}, \qquad (2.1)$$

where $C$ is a constant dependent on the jet energy which fixes the average multiplicity, and $\beta$ is the velocity of the antiproton [12][13]. Interpreted as the differential probability that a quark-jet pair will hadronize to produce a $p\bar{p}$ pair in the final state, such an expression is precisely (twice) the fragmentation function for $q \to \bar{p} + X$. Unfortunately, we need such a parameterization for gluon jets, and data is not available. However, there is some theoretical work which can help to relate the fragmentation of quark and gluon jets. It can be shown that gluon jets should have higher average multiplicities, in the asymptotic energy regime [14]. Also, successful Monte Carlo models confirm the expectation of higher average multiplicities for gluon jets [15]. There is also some expectation that gluon jets give rise to softer hadron spectra [10][16]. However, this effect is not large, and we will choose the fragmentation functions for gluon jets to have the same form as for quark jets, corrected only for a larger average multiplicity by choice of a different normalization constant. The effect of this prescription will be a slight underestimation of the low-energy antiproton spectrum, and this is conservative for our purposes. Furthermore, some fragmentation models can successfully reproduce the observations assuming no difference between gluon- and quark-jet fragmentation [17], and thus we do not consider this prescription to be the leading source of error in our estimate.

Our fit to the data for $e^+e^- \to p\bar{p}X$, as reproduced in Ref. [12] and Ref. [15], gives a quark-jet fragmentation function of the form

$$D_{\bar{p}}(z) = C\beta\left[\exp(-25z) + 0.1\exp(-5.3z)\right]. \qquad (2.2)$$

Note that this data includes contamination from the untagged $b \to \bar{p}X$ fragmentation processes which are expected to have a somewhat different functional form; however, this effect is small.

In order to fix the constant $C$, and thus normalize the gluon fragmentation function, we make use of the fact that the average multiplicity of charged particles in jets follows the approximate formula [12][18]

$$\langle n_{\text{ch}} \rangle = a + b \exp\left[c \left(\ln \frac{s}{Q_0^2}\right)^{1/2}\right]. \qquad (2.3)$$

We find that $a \simeq 2.8$, $b \simeq 0.05$, $c \simeq 2.2$, and $Q_0 \simeq 1$ GeV provides an adequate description of the multiplicity in the Webber model, as illustrated in Figure 10 of Ref. [15], in the jet energy regime in which we are interested. This gives the "per event" multiplicity rather than the "per jet" multiplicity, and thus the factor of two which counts the number of jets is implicit. In addition, we have included the jet multiplicity data for jet events on the $Z^0$ peak [19][11].

Data on exclusive quark-jet multiplicities [20], reproduced in Figure 9 of Ref. [15], together with the Webber model, indicate that the $p\bar{p}$ multiplicity is an essentially constant fraction of the total charged-particle multiplicity over the jet energy range of interest to us, $\langle n_{p\bar{p}} \rangle \simeq 0.06 \langle n_{\text{ch}} \rangle$, with an accuracy of about twenty-five percent. Note also that the production of antineutrons which decay in flight gives a factor of two increase, which we choose to include in the normalization constant rather than display explicitly.

Finally, we must include the contributions to the antiproton spectrum from fragmentation of the $b\bar{b}$ and $c\bar{c}$ final states of neutralino annihilation, which have been discussed previously in the literature [8][9]. We will use the $b, c \to \bar{p}X$ fragmentation functions, $D_{\bar{p}}^{\bar{q}q}(z)$, of Ref. [9]. The factor of two due to antineutron production is included in the normalization of these functions.

The gluon fragmentation function has been normalized by $\int D_{\bar{p}}^{gg}(z)\,dz = \langle n_{p\bar{p}}\rangle_g$, and the normalization of the heavy quark fragmentation functions is included in the parameterization of Ref. [9]. As discussed in Ref. [9], the normalization of the fragmentation functions must be evolved for jet energies away from 10 GeV, where the fit parameters are quoted; this evolution is numerically significant.

Defining the antiproton source function, $q_{\bar{p}}$, as the differential number of produced antiprotons per unit volume, per time interval, per energy interval, we write

$$q_{\bar{p}}(E) = q_{\bar{p}}^{\tilde\chi\tilde\chi\to gg}(E) + q_{\bar{p}}^{\tilde\chi\tilde\chi\to q\bar q}(E), \qquad (2.4)$$

where $q\bar{q}$ indicates the heavy quark final states, $b\bar{b}$ and $c\bar{c}$. At the small velocities typical of WIMPs in the halo, annihilation of neutralinos to light quarks is negligible. It is a simple matter to write the source functions for antiprotons produced in neutralino annihilation in terms of the fragmentation functions and the annihilation cross sections

$$q_{\bar{p}}^{\chi\chi\to gg}(E_{\bar{p}}) = (\sigma v)_{\chi\chi\to gg}\,\langle n_{\tilde\chi}^2\rangle\,\frac{1}{m_\chi}\,D_{\bar{p}}^{gg}(z), \qquad (2.5)$$

$$q_{\bar{p}}^{\chi\chi\to q\bar q}(E_{\bar{p}}) = (\sigma v)_{\chi\chi\to q\bar q}\,\langle n_{\tilde\chi}^2\rangle\,\frac{1}{m_\chi}\,D_{\bar{p}}^{\bar q q}(z), \qquad (2.6)$$

where $z = E_{\bar{p}}/m_\chi$ is the fraction of the jet energy carried by the antiproton, and $\langle n_{\tilde\chi}^2\rangle$ is the volume average of the square of the number density of neutralinos in the galactic halo.

Locally, $n_{\tilde\chi}^{\rm local} = 0.4\,(m_{\tilde\chi}/{\rm GeV})^{-1}\rho_{0.4}\,{\rm cm}^{-3}$, where $\rho_{0.4}$ is the local halo density in units of 0.4 GeV cm$^{-3}$. However, if there is a bulge population of WIMPs and the halo core radius $r_{\rm core}$ is small compared with $R \simeq 8 - 10$ kpc, our distance from the center of the galaxy [21], then there can be a significant enhancement in the volume average of the square of the number density [22],

$$\langle n_{\tilde\chi}^2\rangle \simeq \frac{\pi R}{12 r_{\rm core}}(n_{\tilde\chi}^{\rm local})^2, \quad \text{for } R \gg r_{\rm core}.$$

Now, for illustration, we consider the simplified $B$-ino model discussed in Ref. [4]. If the neutralino is primarily $B$-ino and is less massive than the top quark, then annihilation into gluons can be significant. In addition, there are some theoretical arguments that suggest that the neutralino is likely to be a $B$-ino in this mass range [23]. The neutralino relic abundance $\Omega_{\tilde\chi}$ (in units of the critical density) is determined by the cross section for annihilation of two neutralinos into all lighter particles. The thermally averaged cross section for annihilation of two neutralinos, at a temperature $T$ in the early Universe, is generally written

$$\langle\sigma v\rangle_{\rm ann} \simeq a + b(T/m_{\tilde\chi}). \qquad (2.7)$$

Here, $a$ is the $s$-wave contribution to $\sigma v$ and is the only part which survives in the limit $v \to 0$; $b$ is the $p$-wave contribution. Neutralinos "freeze out" when $T/m_{\tilde\chi} \sim 1/20$. If $a \ll b(T/m_{\tilde\chi})$ when the neutralinos freeze out, then the relic abundance is roughly (see, e.g., Ref. [24])

$$\Omega_{\tilde\chi} h^2 \simeq \left(\frac{b}{8.8 \times 10^{-9}\ {\rm GeV}^{-2}}\right)^{-1}, \qquad (2.8)$$

where $h$ is the Hubble parameter in units of 100 km sec$^{-1}$ Mpc$^{-1}$.

If the neutralino is a pure $B$-ino with mass large compared to the masses of the light fermions, but small compared to the masses of the top quark and the sfermions, $m_b \ll m_{\tilde\chi} \ll m_t, m_{\tilde q}$, then the cross-section coefficients are

$$a \simeq 4.9 \times 10^{-7}\left(\frac{100\ {\rm GeV}}{m_{\tilde f}}\right)^4\ {\rm GeV}^{-2}, \qquad (2.9)$$

and

$$b \simeq 1.2 \times 10^{-6}\left(\frac{m_{\tilde\chi}}{m_{\tilde f}}\right)^2\left(\frac{100\ {\rm GeV}}{m_{\tilde f}}\right)^2\ {\rm GeV}^{-2}. \qquad (2.10)$$

Because of the helicity suppression of majorana-fermion annihilation into light fermions [25], $a$ is suppressed by the squares of the quark and lepton masses, so $a \ll b$. Then, the cosmological abundance of a pure $B$-ino in the given mass range is roughly

$$\Omega_{\tilde\chi} h^2 \simeq 7 \times 10^{-3}\left(\frac{m_{\tilde f}}{m_{\tilde\chi}}\right)^2\left(\frac{m_{\tilde f}}{100\ {\rm GeV}}\right)^2. \qquad (2.11)$$

This implies a relation between the $B$-ino mass and (common) sfermion mass, $m_{\tilde{\chi}} \simeq 0.2 \, (\Omega_{\tilde{\chi}} h^2/0.25)^{-1/2} \, m_{\tilde{f}} \, (m_{\tilde{f}}/100 \text{ GeV})$. Neutralinos can account for the dark matter in galactic halos if the neutralino relic abundance falls in the range $0.03 \lesssim \Omega_{\tilde{\chi}} h^2 \lesssim 1$; if neutralinos are to contribute closure density, then $\Omega_{\tilde{\chi}} h^2 \simeq 0.25$.

At this point we have specified all the parameters necessary to calculate the source spectrum of antiprotons. Recall the approximate cross-sections for the various channels in the pure $B$-ino limit, as discussed in Ref. [4]. Because the two-gluon cross-section depends on the same combination of dimensionful parameters as the relic density, the $B$-ino relic abundance actually fixes the two-gluon cross-section, $\sigma_{gg} v \simeq 4 \times 10^{-12} \, (\Omega_{\tilde{\chi}} h^2/0.25)^{-1} \text{ GeV}^{-2}$. Also, the cross-sections for annihilation to heavy quarks in the non-relativistic limit can be written as

$$\frac{\sigma_{b\bar{b}}}{\sigma_{gg}} \simeq \left(\frac{30 \text{ GeV}}{m_{\tilde{\chi}}}\right)^2,$$
$$\frac{\sigma_{c\bar{c}}}{\sigma_{gg}} \simeq \left(\frac{25 \text{ GeV}}{m_{\tilde{\chi}}}\right)^2. \quad (2.12)$$

Notice that the fraction of annihilations to quarks is relatively small compared to that to all fermions, since the latter is generally dominated by annihilation to $\tau\bar{\tau}$. In addition, in (2.12), we have neglected the effect of running quark masses which would further decrease the contribution of the heavy-quark final states. Thus the gluon final state can be quite important, in comparison to the quark-antiquark final states.

An interesting feature of the gluon contribution to the antiproton production is that the $m_{\tilde{\chi}}$ dependence of the cross-section cancels that which is implicit in the denominator due to the factors of the neutralino number density. In other words, if we assume, as we have done, that the neutralinos make up the dark matter in the galactic halo, with (local) mass density $\rho \simeq 0.4 \text{ GeV/cm}^3$, then their number density in the halo is inversely proportional to their mass. This factor of $m_{\tilde{\chi}}^2$ in the denominator is canceled by a similar factor in the two-gluon cross-section. On the other hand, there is no strong $m_{\tilde{\chi}}$ dependence in the cross section for annihilation into quarks, so the gluon contribution does not decrease as quickly with increasing neutralino mass. Therefore the gluon contribution is relatively more important for larger neutralino masses. Also, there are several other arguments for a relatively important gluon contribution. First, the average charged multiplicity for gluon jets is measurably higher than that for quark jets. Second is a more subtle effect; as an empirical fact, the fragmentation production of a heavy-flavor meson in a heavy-flavor jet is relatively hard [12], compared to the similar process in a light quark jet. On average, this means that less energy is available for baryon production in heavy quark jets. On the contrary, a gluon jet fragments in much the same way as a light quark jet. Clearly the combination of these three effects can be numerically significant for the resulting cosmic-ray antiproton spectrum.

## 3. Results for the Observed Spectrum

Given the above source function $q_{\bar{p}}(E)$ for antiprotons injected in the galaxy, the spectrum at the boundary of the heliosphere, $\phi_{\bar{p}}^{\text{IS}}(E)$, is determined by a model for cosmic-ray propagation through the interstellar medium. The simplest model for interstellar propagation, which we choose, is the leaky-box model [7]. (An alternative which would increase the flux of background antiprotons, the closed-galaxy model [26], seems to be in conflict with recent measurements of the low-energy cosmic-ray $^3\text{He}/^4\text{He}$ ratio [27].) The distortion of the particle spectrum due to propagation in the leaky-box model is described by a (possibly energy dependent) escape time, $\tau_{\text{esc}}(E_{\bar{p}})$, and we write

$$\phi_{\bar{p}}^{\text{IS}}(E_{\bar{p}}) = \frac{1}{4\pi} c \tau_{\text{esc}}(E_{\bar{p}}) q_{\bar{p}}(E_{\bar{p}}). \quad (3.1)$$

The diffusion length contains the effects of galactic escape as well as depletion of the spectrum by interactions, which can be important for less energetic particles. We use $\tau_{\text{esc}} = 10^8$ yrs [28], although we note that estimates for this value vary by roughly an order of magnitude.

To obtain the observed spectrum, $\phi_{\bar{p}}^{\mathrm{obs}}$, the interstellar spectrum must be further corrected for solar modulation. Following Ref. [9], we use the solar modulation model of Perko [29]. The model of Ref. [29] transforms the interstellar spectrum to the observed spectrum by

$$\phi_{\bar{p}}^{\mathrm{obs}}(E) = \phi_{\bar{p}}^{\mathrm{IS}}(E'), \qquad (3.2)$$

where [9][29]

$$E'(E) = \begin{cases} E + \Delta E, & \text{for } p > p_c, \\ p_c \ln \frac{p+E}{p_c+E_c} + E_c + \Delta E, & \text{for } p < p_c. \end{cases} \qquad (3.3)$$

Here $p_c = 1.015$ GeV, $p = (E^2 - m_p^2)^{1/2}$, where $m_p$ is the proton mass, and $\Delta E$ is an energy shift which depends on the phase in the solar cycle.

The results for the modulated cosmic-ray antiproton spectra at solar minimum, expressed as a fraction of the observed cosmic-ray proton spectrum, are displayed in Fig. 1. We took the interstellar cosmic-ray proton spectrum to be $\phi_p = 1.93 \times 10^4 \beta (E/\,\mathrm{GeV})^{-2.7}$ m$^{-2}$ s$^{-1}$ sr$^{-1}$ GeV$^{-1}$ [31], and modulated it in the same way as the antiprotons. The upper (lower) solid curve is the spectrum expected if a 30-GeV (60-GeV) $B$-ino populates the halo. Here we have made the simplest assumptions; we have taken the relic density of WIMPs to be unity, and we have assumed $\langle n_{\tilde{\chi}}^2 \rangle = (n_{\tilde{\chi}}^{\mathrm{local}})^2$. The upper (lower) dashed curve is that which would have been obtained by ignoring the contribution from the two-gluon final state for a 30-GeV (60-GeV) $B$-ino. The dotted line is the "background" flux of cosmic-ray antiprotons expected from spallation of primary cosmic-rays in the leaky-box model [26] suitably modulated at solar minimum. Also shown is the current observational limit of $\bar{p}/p \lesssim 2 \times 10^{-5}$ in the kinetic-energy range 100-1500 MeV [30]. As seen, the two-gluon final state can be the dominant source of antiprotons if the neutralino is primarily $B$-ino. In addition, Fig. 1 illustrates that WIMP annihilation could plausibly provide a low-energy cosmic-ray antiproton signal larger than that expected from background, yet small enough to have evaded detection thus far.

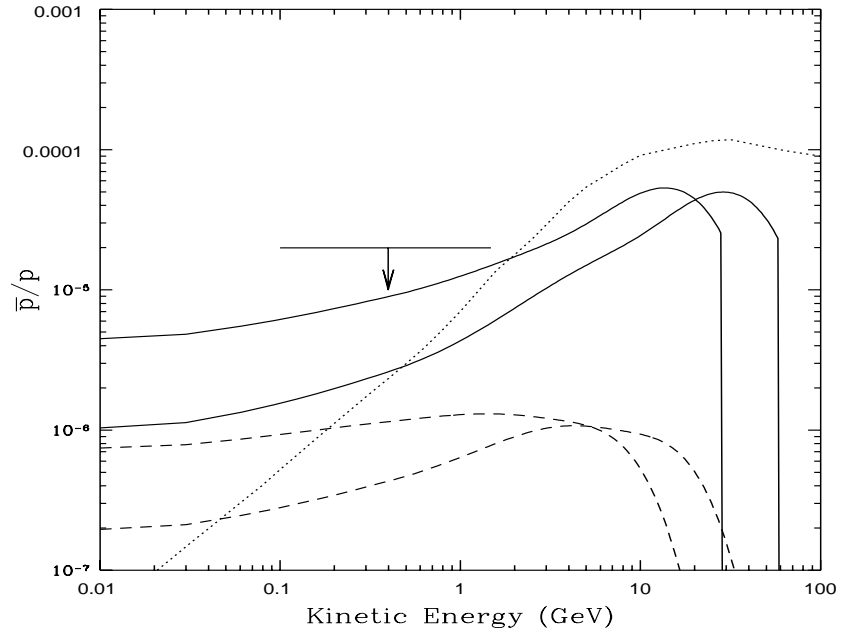

Fig. 1 Observed antiproton/proton ratio as a function of kinetic energy. The dotted curve is the expected solar-modulated flux of "background" antiprotons from standard propagation. The upper (lower) solid curve is the spectrum predicted if a 30-GeV (60-GeV) $B$-ino populates the galactic halo. The upper (lower) dashed curve is the spectrum that would be predicted for a 30-GeV (60-GeV) $B$-ino if we had ignored the contribution of the two-gluon final state. Also shown are the current observational upper limits [30].

4. Discussion

The estimate presented here contains some uncertainties from the physics of hadronization as well as significant uncertainties from the physics of low-energy cosmic-ray antiproton propagation. By considering the source spectrum in some detail we hope that we have illustrated the nature of our results divorced from the latter uncertainties. It would seem difficult to use non-observation of such low-energy antiprotons to constrain dark-matter candidates. Unfortunately, detailed improvements to the calculations are not easily implemented. For instance, many alternate models of cosmic-ray propagation can result in enhancements of the low-energy antiproton flux that might resemble the proposed signal from WIMP annihilation [7][26]. Effects attributable to these models must be separated from any possible new-physics contribution such as neutralino annihilation. Hopefully, improved cosmic-ray data will help discriminate between the various models for secondary antiprotons.

We should also point out that although the WIMP-induced cosmic-ray antiproton spectrum might indeed be observable, such a signal is by no means generic to models where WIMPS account for the dark matter in our halo. There is a relatively limited range of masses and cross sections for which the resulting relic density is large enough and for which the antiproton production cross section is high enough to produce an observable effect. Contrariwise, there are a number of reasonable alterations to the models considered here that could lead to a larger antiproton flux. Once again, we emphasize that the antiproton signal is likely to be enhanced for models where the energetic-neutrino signal from WIMP annihilation in the Sun or Earth is likely to be depleted. Therefore, the search for low-energy cosmic-ray antiprotons provides an important complement to energetic-neutrino searches in the effort to infer the existence of particle dark matter.

## 5. Acknowledgments

We thank G. Tarle, H. Frisch and A. Yagil for helpful discussions and suggestions. M.K. acknowledges the hospitality of the Center for Particle Astrophysics at the University of California at Berkeley and the Aspen Center for Physics where part of this work was completed. M.K. was supported by the Texas National Research Laboratory Commission, and by the U.S. Department of Energy under contract DEFG02-90-ER 40542. G.J. was supported by the U.S. D.O.E. under contract DEFG02-90-ER 40560 at the Enrico Fermi Institute, and under contract DEFG02-85-ER 40231 at Syracuse University.

## References

[1] For recent reviews of dark matter and its detection, see *Proceedings of the ESO-CERN Topical Workshop on LEP and the Early Universe*, ed. J. Ellis, P. Salati, and P. Shaver (CERN preprint TH.5709/90), V. Trimble, *Ann. Rev. Astron. Astrophys.* **25**, 425 (1989); J. R. Primack, B. Sadoulet, and D. Seckel, *Ann. Rev. Nucl. Part. Sci.* **B38**, 751 (1988); *Dark Matter in the Universe*, eds. J. Kormendy and G. Knapp (Reidel, Dordrecht, 1989).

[2] H. E. Haber and G. L. Kane, *Phys. Rep.* **117**, 75 (1985).

[3] J. Ellis, J. S. Hagelin, D. V. Nanopoulos, K. A. Olive, and M. Srednicki, *Nucl. Phys.* **B238**, 453 (1984); K. Griest, M. Kamionkowski, and M. S. Turner, *Phys. Rev. D* **41**, 3565 (1990); M. Drees and M. M. Nojiri, *Phys. Rev. D* **47**, 376 (1993); K. A. Olive and M. Srednicki, *Phys. Lett.* **B230**, 78 (1989); K. A. Olive and M. Srednicki, *Nucl. Phys.* **B355**, 208 (1991); J. McDonald, K. A. Olive, and M. Srednicki, *Phys. Lett. B.* **283**, 80 (1992); K. Griest, *Phys. Rev.* **D38**, 2357 (1988); FERMILAB-Pub-89/139-A (E); *Phys. Rev. Lett.* **61**, 666 (1988).

[4] M. Drees, G. Jungman, M. Kamionkowski, and M. Nojiri, "Neutralino Annihilation to Gluons," submitted to *Phys. Rev. D*.

[5] W. H. Press and D. N. Spergel, *Astrophys. J.* **296**, 679 (1985); A. Gould, *Astrophys. J.* **321**, 571 (1987); A. Gould, *Astrophys. J.* **388**, 338 (1991); J. Silk, K. Olive, and M. Srednicki, *Phys. Rev. Lett.* **55**, 257 (1985); T. Gaisser, G. Steigman, and S. Tilav, *Phys. Rev. D* **34**, 2206 (1986); J. Hagelin, K. Ng, and K. A. Olive, *Phys. Lett. B* **180**, 375 (1987); M. Srednicki, K. Olive, and J. Silk, *Nucl. Phys.* **B279**, 804 (1987); K. Ng, K. A. Olive, and M. Srednicki, *Phys. Lett. B* **188**, 138 (1987); K. A. Olive and M. Srednicki, *Phys. Lett. B* **205**, 553 (1988); L. Krauss, M. Srednicki, and F. Wilczek, *Phys. Rev. D* **33**, 2079 (1986); K. Freese, *Phys. Lett. B* **167**, 295 (1986); F. Halzen, T. Stelzer, and M. Kamionkowski, *Phys. Rev. D* **45**, 4439 (1992); G. F. Giudice and E. Roulet, *Nucl. Phys.* **B316**, 429 (1989).; G. Gelmini, P. Gondolo, and E. Roulet, *Nucl. Phys.* **B351**, 623 (1991); M. Kamionkowski, *Phys. Rev. D* **44**, 3021 (1991); S. Ritz and D. Seckel, *Nucl. Phys.* **B304**, 877 (1988).

[6] J. Silk and M. Srednicki, *Phys. Rev. Lett.* **53**, 624 (1984); J. Ellis *et al.*, *Phys. Lett. B* **214**, 403 (1989); F. Stecker, S. Rudaz, and T. Walsh, *Phys. Rev. Lett.* **55**, 2622 (1985); F. Stecker and A. Tylka, *Astrophys. J.* **336**,

L51 (1989).

[7] T. Gaisser, *Cosmic Rays and Particle Physics*, (Cambridge University Press, Cambridge, 1990).

[8] F. Stecker, S. Rudaz, and T. Walsh, *Phys. Rev. Lett.* **55**, 2622 (1985).

[9] J. Ellis *et al.*, *Phys. Lett. B* **214**, 403 (1988).

[10] R. Field, *Applications of Perturbative QCD* (Addison-Wesley, Redwood City, 1989).

[11] DELPHI collaboration, P. Abreu *et al.*, *Z. Phys. C* **56**, 63 (1992); OPAL collaboration, M.Z. Akrawy *et al.*, *Phys. Lett. B* **247**, 617 (1990);

[12] P. Mattig, *Phys. Rep.* **177**, 141 (1989) and references therein.

[13] B. Naroska, *Phys. Rep.* **148**, 67 (1987).

[14] S.J. Brodsky and J.F. Gunion, *Phys. Rev. Lett.* **37**, 402 (1976).

[15] B.R. Webber, *Nucl. Phys.* **B238**, 492 (1984).

[16] R. Field and R. Feynman, *Nucl. Phys* **B136**, 1 (1978).

[17] P. Mazzanti and R. Odorico, *Nucl. Phys.* **B370**, 23 (1992).

[18] A.H. Mueller, *Phys. Lett. B* **104**, 161 (1981).

[19] L3 Collaboration, B. Adeva *et al.*, *Z. Phys. C* **55**, 39 (1992).

[20] M. Althoff *et al.*, *Z. Phys. C* **17**, 5 (1983).

[21] J. Silk and H. Bloemen, *Astrophys. J.* **313**, L47 (1987); M. S. Turner, *Phys. Rev. D* **34**, 1921 (1986).

[22] M. Kamionkowski and M. S. Turner, *Phys. Rev. D* **43**, 1774 (1991).

[23] L. Roszkowski, *Phys. Lett. B* **262**, 59 (1991).

[24] E. W. Kolb and M. S. Turner, *The Early Universe* (Addison-Wesley, Redwood City, 1990).

[25] H. Goldberg, *Phys. Rev. Lett.* **50**, 1419 (1983).

[26] R.J. Protheroe, *Astrophys. J.* **251**, 387 (1981).

[27] J. J. Beatty *et al.*, *Astrophys. J.* **413**, 268 (1993).

[28] V. L. Ginzburg and V. S. Ptuskin, *Rev. Mod. Phys.* **48**, 161 (1976).

[29] J.S. Perko, *Astron. Astrophys.* **184**, 119 (1987).

[30] S. P. Ahlen *et al.*, *Phys. Rev. D* **61**, 145 (1988); M. Salamon *et al.*, *Astrophys. J.* **349**, 78 (1990); R. E. Streitmatter *et al.*, *Proceedings of the 21st ICRC*, OG 7.3-2, Adelaide **3**, 277 (1990).

[31] M. J. Ryan, J. F. Ormes, V. K. Balasubrahmanyan, *Phys. Rev. Lett.* **28**, 985 (1972).